\documentstyle[pre,aps,epsf]{revtex}

\newcommand{\be}{\begin{equation}}
\newcommand{\ee}{\end{equation}}
\begin{document}
\draft
\title{Equation of state of fully ionized electron-ion 
plasmas\thanks{Scheduled to 
{\it Phys.\ Rev.\/} E {\bf 58}, no.\,4 (October 1998)}}
\author{Gilles Chabrier}
\address{Centre de Recherche Astronomique de Lyon (UMR CNRS \# 5574),\\
     Ecole Normale Sup\'erieure de Lyon,
     69364 Lyon Cedex 07, France}
\author{Alexander Y.~Potekhin}
\address{Ioffe Physical-Technical Institute,
     194021 St.-Petersburg, Russia}

\date{Received 14 April 1998}
\maketitle

\begin{abstract}
Thermodynamic quantities of Coulomb plasmas
consisting of point-like ions immersed in a
compressible, polarizable electron background
are calculated for ion charges $Z=$ 1 to 26 and
for a wide domain of plasma parameters
ranging from the Debye-H\"uckel limit
to the crystallization point 
and from the region of nondegenerate to fully degenerate
nonrelativistic or relativistic electrons.
The calculations are based on the linear-response theory 
for the electron-ion interaction, including the
local-field corrections in the electronic dielectric function. 
The thermodynamic quantities are calculated 
in the framework of the $N$-body hypernetted-chain equations
and fitted by analytic expressions.
We present also accurate analytic approximations
for the free energy of the ideal electron gas
at arbitrary degeneracy and relativity
and for the excess free energy of the one-component plasma
of ions (OCP) derived from Monte Carlo simulations.
 The extension to multi-ionic mixtures
is discussed within the framework of the linear mixing rule. 
These formulae
provide a completely analytic, accurate description
of the thermodynamic quantities of 
fully ionized electron-ion Coulomb plasmas, 
a useful tool for various applications
from liquid state theory to dense stellar matter.
\end{abstract}



\section{Introduction}
Electron-ion plasmas (EIP)
consisting of different species of point-like 
ions (charge $Z_ie$, mass $m_i=A_i$ a.m.u.) 
and electrons ($-e$, $m_e$)
are encountered in numerous physical 
and astrophysical situations like, 
e.g., inertially confined laboratory plasmas, 
liquid metals, stellar and planetary
interiors, supernova explosions, etc. \cite{EOS94}.
Full ionization is reached either at high
temperatures $T$ and low densities $\rho$
(thermal ionization) or
at high enough densities $\rho$ (pressure ionization).
Even when these conditions are not satisfied, 
the approximation of full ionization is useful 
for calculations
in the mean-ion approximation,
in which the mean ion charge
corresponds to its partial ionization stage. 
On the other hand, the free energy of fully ionized EIP provides 
the reference system for 
models aimed at describing the thermodynamic properties 
of partially ionized plasmas \cite{SCVH}. In this paper, we present 
a completely analytic model for the free energy 
of EIP, based on detailed numerical calculations 
for different ionic species $Z$ over a wide 
range of density and temperature. 
We first focus on the two-component plasma (TCP), 
consisting of electrons and a single species of ions. 
Extension to ionic mixtures is considered in Sect.~VI.

The Coulomb plasmas can be characterized 
by the electron coupling parameter $\Gamma_e$ 
and the density parameter $r_s$, 
\be 
    \Gamma_e = \beta e^2/ a_e, \qquad  
    r_s = a_e/a_B, 
\ee 
where $\beta=(k_B T)^{-1}$ 
is the inverse thermodynamic temperature, 
$k_B$ is the Boltzmann constant,  
$a_e=(\frac43\pi n_e)^{-1/3}$ 
measures the mean inter-electron distance, 
$n_e$ is the electron number density,
and $a_B=\hbar^2/(m_e e^2)$ is the Bohr radius.
These parameters can be evaluated as
$\Gamma_e \approx (2.693\times10^5 \,\mbox{K}/ T) \, n_{24}^{1/3}
=(23.2\,\mbox{eV}/k_B T)\, n_{24}^{1/3}$
and
$
   r_s\approx 1.172 \,n_{24}^{-1/3},
$
where $n_{24} \equiv n_e/10^{24}$ cm$^{-3}
\approx(\rho/1.6605{\rm~g~cm}^{-3})\langle Z\rangle/\langle A\rangle$.
Here and hereafter, 
$\langle X\rangle=\sum_i n_i X_i \,/ \sum_i n_i$ denotes 
the average over all ions and $n_i$ the number density of ions
of $i$th species. 

The ion coupling parameter of the TCP is
\be
   \Gamma_i = \beta\, (Ze)^2/ a_i = \Gamma_e\, Z^{5/3},
\ee 
where $a_i$ is the mean interionic distance 
($a_i=a_e Z^{1/3}$ due to the electroneutrality condition $n_e=n_i Z$).
In multicomponent plasmas (MCP), it may be useful 
to define
$
   \Gamma_i = \Gamma_e \langle Z^{5/3} \rangle.
$

The degeneracy parameter $\theta$ and the relativity parameter $x$ are
defined respectively as:
\be
    \theta = T/ T_F,
\qquad 
    x=p_F/(m_e c),
\ee 
where $T_F=(m_e c^2/k_B)\,[\sqrt{1+x^2}-1]
\approx (5.93\times 10^9\,\mbox{K})\,[\sqrt{1+x^2}-1]$ 
is the Fermi temperature,
$c$ is the speed of light, and $p_F=\hbar(3\pi^2 n_e)^{1/3}$ 
is the zero-temperature Fermi momentum of electrons. 
To estimate $\theta$ and $x$, it is useful to note that
\begin{eqnarray}
&&   x =\left( {9 \pi \over 4} \right)^{1/3} {\alpha \over r_s}
      \approx {0.014\over r_s}\approx
       \left( {\langle Z \rangle\over\langle A \rangle}
       {\rho  \over 10^6\mbox{~g~cm}^{-3}}
       \right)^{1/3},
\label{estim-x}
\\&&
   \theta={\alpha^2 (\Gamma_e r_s)^{-1}\over
         \sqrt{1+x^2}-1 },
     \quad \theta \approx 0.543\, {r_s\over\Gamma_e} 
        \quad \mbox{at~} x\ll1,
\label{estim-alpha}
\end{eqnarray}
where $\alpha = 1/137.036$ is the fine-structure constant.

Various asymptotic expansions, 
interpolation formulae 
and large tables have been derived over the past for 
the thermodynamic functions of free fermions
(see Refs.\cite{Blinnikov,Miralles} and references therein).
In this paper, we present analytic 
expressions for the thermodynamic quantities of free fermions 
for arbitrary degeneracy and relativity, $\theta$ and $x$.
Secondly, we propose simple and accurate analytic approximations 
for the {\em nonideal\/} internal and free energies of the 
classical one-component plasma (OCP), which take into account the most recent
hypernetted-chain (HNC) and Monte Carlo (MC) calculations 
by DeWitt et al.\cite{DWSC} (DWSC)
in the strong-coupling regime.
Third, we consider the electron {\em screening\/}
effects on the thermodynamic properties of the TCP.
We employ a computational HNC scheme based 
on the linear screening theory
with local field corrections, taking into account 
the finite temperature (finite $\theta$) effects.
The numerical calculations have been performed over
a wide range of $Z$, $\Gamma_i$, and $r_s$, 
and interpolated by a simple analytic formula, 
which recovers the  
Debye-H\"uckel (DH) limit for the TCP at $\Gamma_i\ll 1$
and the Thomas-Fermi limit at large $\Gamma_i$ and $Z$. 
\section{Summary of the model}
\label{sect-summary}
Consider the Helmholtz 
free energy $F$, internal energy $U$, 
and pressure $P$ of a TCP of $N_i$ ions and $N_e$ electrons
in the volume $V$.
The total free energy $F_{\rm tot}$
can be written as the sum of three terms, 
\be 
   F_{\rm tot} = 
   F_{\rm id}^{(i)} + F_{\rm id}^{(e)} 
   + F_{\rm ex}, 
\label{f-tot}
\ee
where $F_{\rm id}^{(i,e)}$ 
denote the ideal free energy
of ions and electrons, respectively, 
and $F_{\rm ex}$ 
is the {\it excess\/} free energy arising from 
interactions.

In this paper, we restrict ourselves to conditions where 
the ions behave classically, which is the case 
in most astrophysical situations.
Quantum corrections for ions which can be important
in the ultradense matter of white dwarf interiors, 
neutron stars, and supernova cores
have been considered, e.g., in Refs.\cite{C93,JonesCeperley}.
Thus $F_{\rm id}^{(i)}$ is given by the Maxwell-Boltzmann expression.
For $F_{\rm id}^{(e)}$, we use the well-known
expressions of the thermodynamic functions of 
the perfect gas of fermions 
(which may be 
degenerate and relativistic)
through the generalized Fermi-Dirac integrals.

To calculate $F_{\rm ex}$, we follow the model developed by Chabrier 
\cite{Chabrier90} for fully ionized EIPs.
As long as the ion-electron interaction 
is weak compared to the kinetic energy of the electrons, 
$Ze^2/a_e\ll k_B T_F$, 
this interaction can be treated within the linear screening theory. 
Under these conditions, the exact Hamiltonian of the TCP 
can be separated out exactly into a Hamiltonian 
for the electron-screened ionic fluid 
and a Hamiltonian for a {\it rigid} electron background, 
the so-called ``jellium" Hamiltonian $H_e$ 
\cite{GalamHansen,AshcroftStroud}:
\be 
H=H^{\rm eff}\,+\,H_e
  \label{hamil}
\ee 
with
\be 
H^{\rm eff}=K_i\,+\,\frac{1}{2V}\sum_{{\bf k} \ne 0}\frac{4\pi (Ze)^2}{k^2}
\left[\frac {\rho_{\bf k}\rho_{\bf k}^\ast}{\epsilon(k)}-N_i\right]
  \label{heff}
\ee 
where $K_i$ is the ionic kinetic (translational) term, 
$\rho_{\bf k}$ is the Fourier component 
of the ionic microscopic density and $\epsilon(k)$ 
is the static screening function of the electron fluid 
to be discussed below. The 
Hamiltonian $H^{\rm eff}$
characterizes the electron-screened ion fluid with the interparticle 
potential whose Fourier transform is
\be 
V^{\rm eff}(k)=\frac{4\pi (Ze)^2}{k^2\epsilon(k)},
  \label{veff}
\ee 
which is the sum of the bare ionic potential 
and the induced polarization potential.

The ion-ion ($ii$) and the ion-electron ($ie$) Coulomb 
interactions can thus be separated from the
exchange-correlation
contribution in the electron fluid ($ee$). 
The excess part of the free energy (\ref{f-tot}) can then be 
written as $F_{\rm ex}=F_{ee}+F_{ii}+F_{ie}$; the quantities
labeled $ie$ will be referred to as {\it electron-screening}
quantities.
It is convenient to consider dimensionless quantities
$f_{ee}\equiv\beta F_{ee}/N_e$ 
and $f_{ii,ie}\equiv\beta F_{ii,ie}/N_i$.
Then
\be 
   f_{\rm ex} = 
   x_e f_{ee} +
   x_i (f_{ii} + f_{ie}),
\label{f-ex}
\ee 
where $x_{i,e}\equiv N_{i,e}/N$ denote 
the number fraction of ions and electrons, respectively, 
and $N=N_i+N_e$ is the total number of particles.
In the same way we define $u_{ee}\equiv\beta U_{ee}/N_e$
and $u_{ii,ie}\equiv\beta U_{ii,ie}/N_i$.
The excess free energy can be obtained from 
the internal energy by integration:
\be
   f_{\rm ex}(\Gamma,r_s)=
\int_0^\Gamma {u_{\rm ex}(\Gamma,r_s)\over\Gamma}\, {\rm d} \Gamma .
\label{int_fren}
\ee

For $f_{ee}$, we have adopted 
the interpolation formula of Ichimaru et al.\cite{IIT} (hereafter IIT), 
consistent with numerical results obtained 
by different authors. 

For $f_{ii}$, which corresponds to the well-known 
OCP model, 
that implies the rigid electron background ($\epsilon(k)=1$), 
we present an analytic interpolation 
between the MC results 
\cite{DWSC}
at $\Gamma_i\ge 1$ and the DH limit 
and Abe correction at $\Gamma_i\lesssim0.1$. 

The ion-electron interactions are calculated 
numerically as in Ref.\cite{Chabrier90}. 
In this approach,
the bare Coulomb potential in the expression for the electrostatic
energy is replaced by the potential statistically 
screened by the electrons (\ref{veff}),
and the HNC approximation is used to 
calculate the thermodynamic functions of the system.
This model, originally applied to nonrelativistic hydrogen plasmas,
is now extended to the case of arbitrary $Z$ and $x$.
In the nonrelativistic case ($x\ll 1$),
the dielectric function $\epsilon(k)$ 
is the finite-temperature Lindhard function 
modified with the local field 
correction arising from electron correlation effects. 

At very high density, $x\gtrsim1$, the electrons 
become relativistic. At such densities, the electron
correlation effects are completely negligible. 
The finite-temperature effects ($\theta\neq 0$)
may give an appreciable contribution to the screening
part of the free energy $f_{ie}$ only at extremely high temperatures,
where the nonideality of the gas has no significance.
Thus we use the Jancovici \cite{Janco} 
zero-temperature dielectric function in the relativistic regime.

The correlation functions and thermodynamic quantities 
for the electron-screened ionic fluid are obtained within the framework 
of the HNC equations. The validity of the HNC
theory for the Coulomb systems has been assessed 
by several authors by comparison with lengthy MC simulations.
The HNC approximation consists of neglecting the contribution 
of the so-called bridge diagrams, which involves an infinite 
series of multiple integrals, in the $N$-body general 
diagrammatic resummations \cite{HMc}. The long-range part of the
direct correlation function $c(r)$ calculated within 
the HNC approximation is exactly canceled by $-V(r)/kT$, 
so that the pair correlation function $g(r)$ is
of much shorter range than the Coulomb potential $V(r)$ \cite{HMc}. 
This is a required condition for Coulomb systems, because of the 
perfect screening condition. This property of the HNC theory 
makes it particularly suitable for such long-range systems. 
The differences on the free energy, 
the internal energy and the pressure 
are at most of the order of 1\% (see, e.g., Refs.\cite{DWSC,Chabrier90}). 
The difference is due to the lack of bridge functions in the HNC theory. 
\section{Ideal part of the free energy}
The ideal free energy of nonrelativistic classical ions, 
neglecting their spin statistics,
reads\cite{LaLi}: 
\be 
   F_{\rm id}^{(i)} = N_i k_B T \left[\ln(n_i\lambda_i^3)-1\right],
\label{id_i}
\ee 
where $\lambda_i= (2\pi\beta\hbar^2/m_i)^{1/2}$ 
is the thermal wavelength of ions. 
For electrons, we use the identity\cite{LaLi}
\be 
   F_{\rm id}^{(e)} = 
   N_e\mu_{\rm id}^{(e)} - P_{\rm id}^{(e)} V .
\label{id_e}
\ee 
Here, $\mu_{\rm id}^{(e)}$  
is the 
chemical potential (in which we do not include 
the rest energy $m_ec^2$) and $P_{\rm id}^{(e)}$ is the 
pressure of the ideal Fermi gas. 
The pressure and number density, in turn, are functions of 
$\mu$ and $T$:
\begin{eqnarray}
   P_{\rm id}^{(e)} &=& {(2m_e)^{3/2}
  \over 3\pi^2\hbar^3\beta^{5/2} }
   \left( I_{3/2}(\chi,\tau)
   + {\tau\over 2}I_{5/2}(\chi,\tau) \right),
\\
   n_e &=& {\sqrt{2}\,(m_e/\beta)^{3/2} \over \pi^2\hbar^3 }
   \left( I_{1/2}(\chi,\tau)
   + \tau I_{3/2}(\chi,\tau) \right),
\label{n_e}
\end{eqnarray}
where $\tau = (\beta m_e c^2)^{-1} = T/5.93\times 10^9$~K, 
$\chi = \beta\mu_{\rm id}^{(e)}$, and
\be 
   I_\nu(\chi,\tau) \equiv \int_0^\infty
  { x^\nu\,\sqrt{1+\tau x/2}
    \over \exp(x-\chi)+1 } {\rm d}x 
\label{I_nu}
\ee 
is the generalized Fermi-Dirac integral. 

In the limit $\tau\to 0$, the Fermi-Dirac integrals 
reduce to 
the usual nonrelativistic Fermi integrals $I_\nu(\chi)$,
which  
can be calculated using 
the highly accurate Pad\'e approximations 
presented by Antia \cite{Antia}. 
The chemical potential is obtained from the relationship 
\be 
   \chi = X_{1/2}(2\theta^{-3/2}/3),
\label{chi0}
\ee 
where $X_\nu$ is the inverse Fermi integral,
also fitted with high accuracy by Antia \cite{Antia}. 

The accuracy of the nonrelativistic formulae  
decreases rapidly at $T>10^7$~K. 
Blinnikov et al.\cite{Blinnikov} have presented 
a number of approximations and asymptotic 
expansions of the relativistic thermodynamic 
functions of the ideal electron gas. We have 
selected those of their fitting formulae that are most
accurate at low and moderate $\chi$
and supplemented them  
with asymptotic expansions at high $\chi$ 
to obtain an approximation which is accurate 
at any $n_e$ 
for each of the Fermi integrals $I_\nu(\chi,\tau)$ with
$\nu=\frac12,\frac32$, and $\frac52$ :
\begin{eqnarray}
   I_{k+1/2}(\chi,\tau) &=& \sum_{i=1}^5 c_i^{(k)}
   { \sqrt{1+\chi_i^{(k)}\tau/2}   
   \over \exp(-\chi_i^{(k)})+\exp(-\chi) }
\qquad ( \chi\leq 0.6),
\label{fitIa}
\\
   &=& \sum_{i=1}^5 \left[ h_i x_i^k \,
 { \chi^{k+3/2} \sqrt{ 1+\chi x_i\tau/2} 
   \over 1+\exp(\chi x_i - \chi) }
 \,\, + \,\, v_i (\xi_i + \chi)^{k+1/2} 
 \sqrt{1+(\xi_i+\chi)\tau/2} \rule{0em}{3ex}\right]
\nonumber\\
&&
\qquad (0.6 < \chi < 14),
\label{fitIb}
\\
   &=& F_k(\chi,\tau) + {\pi^2\over6} \,\chi^k 
 {k+1/2+(k+1)\chi\tau/2\over R}
\qquad  ( \chi \geq 14) ,
\label{fitIc}
\end{eqnarray}
where $R \equiv \sqrt{\chi(1+\chi\tau/2)}$, 
\begin{eqnarray}
   F_0(\chi,\tau) &=& (\chi+\tau^{-1})R/2 
  - (2\tau)^{-3/2}\ln(1+\tau\chi+\sqrt{2\tau}\,R),
\\
   F_1(\chi,\tau) &=& \left(2 R^3/3 - F_0(\chi,\tau)\right)/\tau,
\\
   F_2(\chi,\tau) &=& \left(2\chi R^3 
  - 5 F_1(\chi,\tau) \right)/(4\tau).
\end{eqnarray}
If $\chi\tau \ll 1$, the functions $F_k(\chi,\tau)$ should be 
replaced by their nonrelativistic limits, 
$\chi^{k+3/2}/(k+3/2)$. 
The constants $c_i^{(k)}$, $\chi_i^{(k)}$, 
$x_i$, $\xi_i$, $h_i$, and $v_i$ are 
adopted from Ref.\cite{Blinnikov} and listed in Table~\ref{tab1}.
The relative error of the approximation (\ref{fitIa})--(\ref{fitIc}) 
does not exceed 0.2\% at $\tau \leq 10^2$ (any $\chi$), 
being typically a few parts in $10^4$. 

\begin{table*}
\caption{Parameters of Eqs.~(\protect\ref{fitIa}) and 
(\protect\ref{fitIb}).
The powers of 10 are given in square brackets}
\label{tab1}
\begin{tabular}{llllll}
  $i$ & 1 & 2 & 3 & 4 & 5 \\
\noalign{\smallskip}
\hline
\noalign{\smallskip}
$c_i^{(0)} $ & $ 0.37045057 $ & $  0.41258437 
     $ & $  9.777982\,[-2] $ & $  5.3734153\,[-3] $ & $  3.8746281\,[-5] $\\ 
$c_i^{(1)} $ & $0.39603109 $ & $  0.69468795 
     $ & $  0.22322760 $ & $  1.5262934\,[-2] $ & $  1.3081939\,[-4] $\\ 
$c_i^{(2)} $ & $0.76934619 $ & $  1.7891437   
     $ & $  0.70754974 $ & $  5.6755672\,[-2] $ & $  5.5571480\,[-4] $\\ 
$\chi_i^{(0)} $ & $ 0.43139881 $ & $  1.7597537 
     $ & $  4.1044654 $ & $  7.7467038 $ & $  13.457678 $ \\ 
$\chi_i^{(1)} $ & $  0.81763176 $ & $  2.4723339 
     $ & $  5.1160061 $ & $  9.0441465 $ & $  15.049882 $\\ 
$\chi_i^{(2)} $ & $  1.2558461 $ & $  3.2070406   
     $ & $  6.1239082 $ & $  10.316126 $ & $  16.597079 $\\ 
$x_i$ & $ 7.265351\,[-2] $ & $  0.2694608 
     $ & $  0.533122 $ & $  0.7868801 $ & $  0.9569313 $\\ 
$\xi_i$ & $ 0.26356032 $ & $  1.4134031   
     $ & $   3.5964258 $ & $  7.0858100 $ & $  12.640801 $\\ 
$h_i$ & $ 3.818735\,[-2] $ & $  0.1256732 
     $ & $  0.1986308 $ & $  0.1976334 $ & $  0.1065420 $\\
$v_i$ & $0.29505869 $ & $  0.32064856 $ & $  7.3915570\,[-2] 
     $ & $  3.6087389\,[-3] $ & $2.3369894\,[-5] $\\
\end{tabular}
\end{table*}

The chemical potential $\mu_{\rm id}^{(e)}$ can be 
obtained numerically from Eq.~(\ref{n_e}), using 
Eqs.~(\ref{fitIa})--(\ref{fitIc}). 
We have constructed also
an analytic fit to $\chi$:
\be 
   \chi = \chi^{\rm nonrel} - 
       {3\over 2} \ln\left[ 1+ \left( {\tau\over 1+\tau/(2\theta)} \right)
          {1+q_1\,\sqrt{\tau}+q_2 q_3\tau \over
                1+q_2\tau } \right]. 
\label{chi}
\ee 
Here $\chi^{\rm nonrel}$ is given 
by the nonrelativistic formula (\ref{chi0}), 
and the coefficients $q_i$ are functions of $\theta$:
\begin{eqnarray}
q_1  &=&  \frac32 ({\rm e}^\theta-1)^{-1},
\nonumber\\
q_2  &=&  12 + 8\theta^{-3/2},
\nonumber\\
   q_3  &=&  {2\over\pi^{1/3}} 
         - {{\rm\,e}^{-\theta}+1.612{\rm\,e}^\theta
      \over 6.192\,\theta^{0.0944}{\rm\,e}^{-\theta}+
  5.535\,\theta^{0.698}{\rm\,e}^{\theta} }.
\nonumber
\end{eqnarray}
The relative error $\delta\chi/\chi$ becomes infinite at $\chi=0$. 
However, since thermodynamic quantities are expressed 
through $\chi$ by virtue of 
thermal averaging of type of Eq.~(\ref{I_nu}), 
a natural measure of the error
is $\delta\chi/\max(|\chi|,1)=\delta\mu/\max(|\mu|,k_B T)$. 
The error thus lies within 0.4\% 
for $\tau > 1$ and is smaller than 0.2\% if $\tau < 1$ (any $\theta$).  
Another measure of the accuracy is the relative difference
between the densities $n_e$ calculated with 
the exact and fitted values of $\mu$.
This difference lies within 0.4\% for $\tau\geq1$
and within 0.1\% for $\tau < 1$.

This accuracy may not be sufficient for 
calculation of temperature derivatives of the electron-gas EOS
(heat capacity, temperature exponent, etc.)
in the regime of strong degeneracy ($\chi\gg1$).
In this case, however, Sommerfeld asymptotic expansions 
for these quantities may be used (see, e.g., Ref.\cite{YS}).

In this paper we do not consider the positrons,
which are efficiently created at $\tau\gtrsim1$
(see Ref.\cite{Blinnikov} for description
of the equilibrium electron-positron plasma).

\section{OCP liquid of classical ions}
\label{sect-ii}
Liquid and solid phases of the OCP have been 
studied extensively 
by various numerical methods, MC simulations or $N$-body 
semianalytic theories like the HNC theory 
(see Refs.\ \cite{BausHansen,IIT} for detailed
reviews).
All the thermodynamic functions of the OCP
of classical ions 
in a uniform (rigid) electron background
can be expressed as functions of the only parameter
$\Gamma_i$.
The melting point of the OCP
corresponds to $\Gamma_i\approx172$, above which it forms
a Coulomb crystal\cite{NNN}.
The most accurate MC and HNC results
for the internal and free energies of the liquid OCP
for $1\leq\Gamma_i\leq160$ have been obtained recently by DWSC \cite{DWSC}
(see references therein for earlier results).
The high precision of the calculations allowed the authors 
to investigate the tiny effects of nonadditivity
of the excess energy of binary ionic mixtures, 
as will be commented in Sect.~VI.

DWSC have also derived a highly accurate analytic fit 
to the MC simulations of the internal energy of the OCP 
in the aforementioned $\Gamma$-range: 
\be 
   u_{ii} = a\Gamma_i + b\Gamma_i^s +c,
\label{dwscfit}
\ee 
with $a =-0.899126$, $b=0.60712$, $c=-0.27998$, 
and $s=0.321308$. 
The maximum relative difference between calculated and fitted 
values reaches 17 parts in $10^5$ at $\Gamma_i=3.1748$.

Equation (\ref{dwscfit}), however, does not apply to
the weak-coupling region $\Gamma_i < 1$.
At very small $\Gamma_i$, the internal energy 
of the OCP must recover the well-known DH expression
$u_{ii} = -(\sqrt{3}/2)\,\Gamma_i^{3/2}$ whereas
at moderately small $\Gamma_i$ this limit must include the 
Abe correction \cite{Abe}:
\be 
   u_{ii} = -{\sqrt{3}\over2}\,\Gamma_i^{3/2}
            -3\Gamma_i^3\,\left[\frac38\ln(3\Gamma_i)
            +{\gamma\over2}-\frac13\right],
\label{abe}
\end{equation}
where $\gamma=0.57721\ldots$ is the Euler's constant.

We represent the internal energy of 
the ionic fluid ($\Gamma_i\lesssim170$) by
a simplified version of the fitting formula
proposed by Hansen\cite{Hansen73,HTV},
\be
   u_{ii} = \Gamma_i^{3/2}
   \left[{A_1\over\sqrt{A_2+\Gamma_i}} 
   + {A_3\over 1+\Gamma_i}\right], 
\label{fitionu}
\ee
where $A_1$ and $A_2$ are fitting parameters,
and $A_3=-\sqrt{3}/2-A_1/\sqrt{A_2}$.
We have found that the minimum relative 
difference between Eq.~(\ref{fitionu}) and the MC results
of DWSC\cite{DWSC} --- smaller than 6 parts in $10^4$ --- 
is obtained with $A_1=-0.9052$ and $A_2=0.6322$.
This accuracy is sufficient for our present study
since it is much better than the available numerical
accuracy of the complementary 
contribution to the internal energy, $u_{ie}$. 
As mentioned in Sect.~\ref{sect-summary},
the HNC calculations of  the {\em sum\/} $u_{ii}+u_{ie}$
ensure an accuracy of the order of 1\%.

\begin{figure}
    \begin{center}
    \leavevmode
    \epsfysize=90mm
    \epsfbox[80 160 520 670]{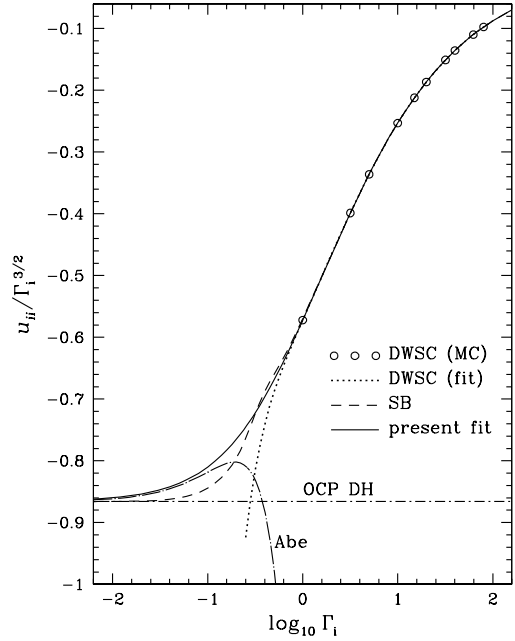}
    \end{center}
\caption{
Comparison of the fit (solid line) given by Eq.~(\protect{\ref{fitionu}}) 
for the OCP internal energy
with the DH and Abe asymptotic expansions at small $\Gamma_i$ (dot-dash lines)
and with the DWSC results\protect\cite{DWSC} 
at $1\leq\Gamma_i\leq160$ (dots and circles).
The dashed curve presents the interpolation of SB\protect\cite{SB}.
}
\label{fig-ii}
\end{figure}

Figure \ref{fig-ii} 
presents a comparison of our interpolation formula (\ref{fitionu})
with the DH-Abe formulae,
the MC results and fit (\ref{dwscfit}) of DWSC\cite{DWSC}, 
and the interpolation proposed 
by Stolzmann and Bl\"ocker 
\cite{SB} (hereafter SB) following Ebeling \cite{Ebeling90}.
Unlike SB, our Eq.~(\ref{fitionu}) accurately 
reproduces Eq.~(\ref{abe})
in the range $\Gamma_i \sim 0.01 - 0.1$
and provides a smoother transition between the 
strong- ($\Gamma_i >1$) and weak- ($\Gamma_i \ll 1$) coupling regimes. 

Using Eqs.~(\ref{fitionu}) and (\ref{int_fren})
we obtain the Helmholtz free energy (cf.\ Ref.\cite{Hansen73}):
\begin{eqnarray}
   f_{ii}(\Gamma_i) &=& A_1\left[\sqrt{\Gamma_i(A_2+\Gamma_i)}
           -A_2\,\ln\left(\sqrt{\Gamma_i/ A_2}
           +\sqrt{1+\Gamma_i/ A_2}\right)\right]
\nonumber\\&&
           + 2A_3\left[\sqrt{\Gamma_i}
           -\arctan\left(\sqrt{\Gamma_i}\right) \right].
\label{fition}
\end{eqnarray}
At $\Gamma_i\geq1$, this formula gives $f_{ii}$ which differs 
from the HNC calculations and fit of DWSC\cite{DWSC}
by no more than 0.8\%.
This difference approximately
coincides with that between the MC and HNC results for $u_{ii}$,
therefore it should be attributed to the lack of the bridge functions
in the HNC approximation (see Sect.~\ref{sect-summary}).
On the other hand,  Eq.~(\ref{fition}) 
recovers the DH-Abe free energy with
an error smaller than 0.6\% at $\Gamma_i < 0.1$.

\section{Electron fluid}
The exchange and correlation effects in electron fluid
were studied by many authors.
For instance, 
Tanaka et al.\cite{Tanaka85} calculated the interaction energy 
of the electron fluid at finite temperature
in the Singwi-Tosi-Land-Sj\"olander 
\cite{STLS} approximation 
and presented a fitting formula that reproduces their 
results as well as various exact limits
with disgressions less than 1\%
(in particular, their formula incorporates
the parametrization of the exchange energy 
by Perrot and Dharma-wardana \cite{Perrot}).
We adopt a modification of this formula given 
by IIT \cite{IIT}. 

The exchange-correlation free energy, $f_{ee}$, is obtained by 
integration from Eq.~(\ref{int_fren}). 
It is important to note that Tanaka et al.\cite{Tanaka85} 
give a fit to the {\it interaction\/} energy of the electron fluid
but not to the thermodynamic {\it internal\/} energy
(the quantities differ at finite $\theta$).
This enabled Tanaka et al. to obtain
$f_{ee}$ by integration of their fitting formula 
over $\Gamma_e$ at constant $\theta$ (the integration 
of the internal energy would have to be performed 
at constant $r_s$). Note also that the results of IIT
are nonrelativistic.

More recently, SB \cite{SB} proposed other parametrizations of
the exchange and correlation free energies.
At moderate $r_s$, comparison of the formulae given by SB and IIT  
reveals only small differences, which 
do not exceed the uncertainty in the various numerical results
found in the literature\cite{Perrot,xc}. 
Unlike IIT, 
SB evaluated the exchange energy at $\theta < 1$ 
in the relativistic case.
On the other hand, the SB fit 
reaches the classical OCP limit at large $r_s$ 
and moderate $\Gamma_e$ with disgressions
up to 4.4\%, while the parametrization of IIT
is several times more accurate in this limit.
We shall use the IIT's formula hereafter.

\section{Electron screening}
\subsection{Numerical calculations}
In order to calculate the screening contribution, 
we have employed the model 
of Ref.\cite{Chabrier90}, outlined in Sect.~\ref{sect-summary}. 
The HNC equations were solved numerically 
for the effective screened 
interionic potential (\ref{veff})
to obtain $f_{ii}+f_{ie}$, $u_{ii}+u_{ie}$, and $P_{ii}+P_{ie}$,
and for the bare Coulomb potential
to obtain $f_{ii}$, $u_{ii}$, and $P_{ii}$. 
The difference represents 
the screening ($ie$)
contribution to the thermodynamic quantities. 

\begin{figure}
    \begin{center}
    \leavevmode
    \epsfysize=85mm
    \epsfbox[30 160 550 660]{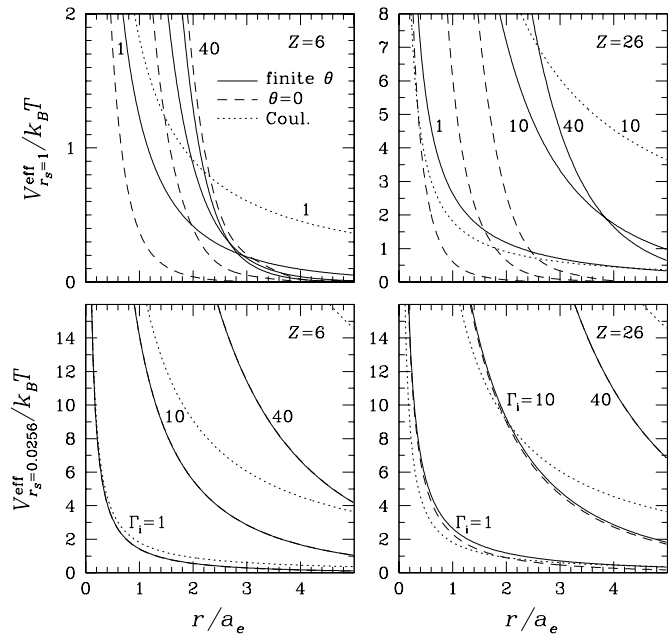}
    \end{center}
\caption{
Effective ion-ion potentials $V^{\rm eff}$
at various approximations
for $Z=6$ (left panels) and $Z=26$ (right),
for two densities, $r_s=1$ (upper panels)
and 0.0256 (lower ones) and 
three values of $\Gamma_i=1$, 10, and 40.
Solid lines represent 
finite-temperature $V^{\rm eff}$
including the local field correction,
and dashes show the zero-temperature RPA approximation
(the dashed and solid lines practically coincide
on the lower left panel). 
Bare Coulomb potential is drawn by dots 
for comparison.
}
\label{fig-Veff}
\end{figure}

The previous numerical results \cite{Chabrier90}
have been obtained for the hydrogen plasma ($Z=1$).
We extend the calculations to different values 
of $Z$ and a larger set of $r_s$.
Figure \ref{fig-Veff} shows the effective
potentials $V^{\rm eff}$ for $Z=6$ and $Z=26$
at several values of $r_s$,
compared with the bare Coulomb potential and
with $V^{\rm eff}$ in the zero-temperature ($\theta=0$) 
RPA approximation (no local field correction).
One can see that the latter approximation works well
at the small value of $r_s=0.0256$ (lower panels),
while it breaks down completely at $r_s=1$ (upper panels).

\begin{figure}
    \begin{center}
    \leavevmode
    \epsfysize=90mm
    \epsfbox[60 150 540 670]{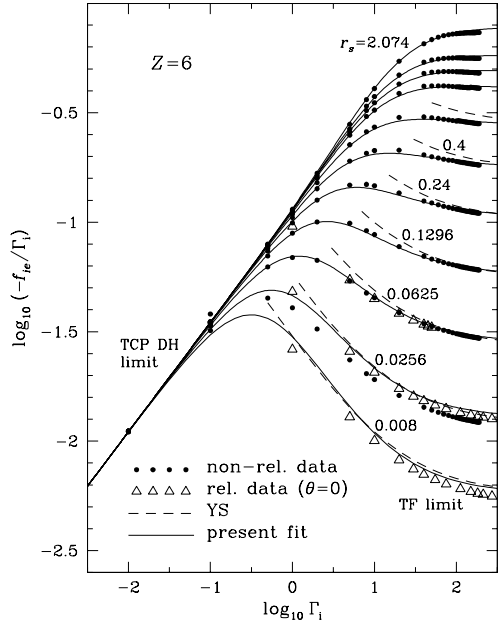}
    \end{center}
\caption{
Nonrelativistic finite-temperature (filled circles)
and 
relativistic zero-temperature (open triangles) calculated values
of the screening part $f_{ie}$ of the free energy of TCP for $Z=6$,
compared with the fit (\protect\ref{fitscr}) (solid lines).
The fit is also compared with 
the approximation of YS\protect\cite{YS} (dashed line),
which is valid at small $r_s$ and large $\Gamma_i$.
}
\label{fig-z6}
\end{figure}

The bulk of the calculations has been performed 
in the nonrelativistic approximation, for 13 ion charges from 
$Z=1$ to $Z=26$ listed in the first column of Table~\ref{tab2}, 
at 10 values of the density 
parameter $r_s$ ranging from $r_s=0.0256$ to $r_s\approx2$, 
and, at each $Z$ and $r_s$, for several tens of values  
of the coupling parameter $\Gamma_i$
which range from
the DH limit at $\Gamma_i = 0.001$ 
to $\Gamma_i \sim 200$. 
As an example, calculated values of  
the normalized screening part of the free energy 
$f_{ie}$ at $Z=6$ are shown 
by filled circles in Fig.~\ref{fig-z6}. 
Note that it is the account of the finite electron temperature
in the dielectric function that allows us to
reach the correct TCP DH limit at low values of $\Gamma_i$ 
(see Ref.\cite{Chabrier90}).

In order to supplement the aforementioned nonrelativistic
data at higher densities, we have also 
performed calculations using the Jancovici \cite{Janco}
zero-temperature dielectric function.
The results are shown in Fig.~\ref{fig-z6} for
$\Gamma_i\geq 1$ and $r_s = 0.0625$ ($x=0.224$),
0.0256 ($x=0.545$), and 0.008 ($x=1.75$).
A comparison for the case of $r_s=0.0625$ confirms 
that the zero-temperature 
approximation works well at small $r_s$ and large $\Gamma$
(where $\theta\ll1$ due to the relation (\ref{estim-alpha}));
this conclusion is corroborated by inspection of Fig.~\ref{fig-Veff}.

Thus the numerical results cover all values of $r_s$ and $\Gamma$
(i.e., $\rho$ and $T$)
that are relevant for liquid EIPs.
At $r_s \gtrsim 1$ and $\Gamma_i\gtrsim 1$,
the formation of bound states sets in.
At $r_s\lesssim10^{-2}$ and $\Gamma_i\lesssim 1$,
the temperature reaches the values 
$T\gtrsim 3\times 10^7 Z^{5/3}$~K,
where the electron screening
effects are completely unimportant.
Finally, at $\Gamma_i\gtrsim 170$, solidification takes place.

\subsection{Analytic formulae}
The calculated values of the screening free energy are fitted
by the following function of $r_s$, $\Gamma_e$, and $Z$:
\be 
   f_{ie} = -\Gamma_e \,
   { c_{DH} \sqrt{\Gamma_e}+
    c_{TF} a \Gamma_e^\nu g_1(r_s) h_1(x)
    \over
     1+\left[ b\,\sqrt{\Gamma_e}+ a g_2(r_s) \Gamma_e^\nu/r_s \right]
    h_2(x) }.
\label{fitscr}
\ee 
The parameter
\be 
   c_{DH} = {Z\over\sqrt{3}}
    \left[(1+Z)^{3/2}-1-Z^{3/2}\right]
\ee 
ensures transition to the DH 
value of the excess free energy of the EIP,
$f_{\rm ex}^{DH}=-Z\,[(1+Z)/3]^{1/2}\,\Gamma_e^{3/2}$
at small $\Gamma_e$. 
The parameter 
\be
   c_{TF} =
   c_\infty Z^{7/3}\left(1-Z^{-1/3}+0.2\,Z^{-1/2}\right) 
\ee
determines 
the screening in the limit
of large $\Gamma_e$ and small $r_s$.
The parameter $c_\infty=(18/175)(12/\pi)^{2/3}=0.2513$ 
is consistent with the Thomas-Fermi
approximation \cite{Salpeter}, 
which becomes exact at small $r_s$ and 
very large $Z$ (cf.\ Ref.\cite{YS}).
The parameters 
\begin{eqnarray}
   a & = & 1.11 \, Z^{0.475},
\nonumber\\
   b &=& 0.2+0.078 \,(\ln Z)^2,
\nonumber\\
   \nu &=& 1.16 + 0.08 \ln Z
\nonumber
\end{eqnarray}
provide a low-order approximation to $f_{ie}$
(with a maximum error up to 30\% at large $Z$ and $r_s\gtrsim1$),
while the functions
\begin{eqnarray}
   g_1(r_s) &=& 1 +
              {0.78\over 21+ \Gamma_e(Z/r_s)^3}
               \left({\Gamma_e\over Z}\right)^{1/2},
\label{g1}\\
   g_2(r_s) &=& 1+{ Z-1 \over 9}
         \left(1+\frac{1}{0.001\, Z^2+2\Gamma_e}\right)
         {r_s^3 \over 1+ 6 \, r_s^{2} }
\label{g2}
\end{eqnarray}
improve the fit at relatively large $r_s$
and reduce the maximum fractional error
in $f_{ie}$ to 4.3\%,
and the root-mean-square (rms) error to $\sim1.5$\%.

The factors $h_1(x)=[1+(v_F/c)^6 Z^{-1/3}]^{-1}$ (where 
$v_F=cx/\sqrt{1+x^2}$ is the electron Fermi velocity) 
and $h_2(x)=(1+x^2)^{-1/2}$ 
are relativistic corrections and may be omitted at $x\ll1$.

Note that $f_{ie}$ constitutes only a part
of the ion excess free energy $f_{ii}+f_{ie}$. 
The fit to this latter quantity is given by the sum 
of Eqs.~(\ref{fition}) and (\ref{fitscr}).
The second and third columns of
Table~\ref{tab2} present the rms and maximum 
relative differences between
the calculated and fitted values of $f_{ii}+f_{ie}$
at each value of $Z$.
The comparison has been done for the set of 
finite-temperature numerical results at $0.1 \leq\Gamma_i\leq 170$
and $0.0625 \leq r_s \leq 2.074$.
The remaining four 
columns of the table present the rms and maximum relative
differences for the ($ii+ie$) internal energy and pressure,
derived from the fits by the use of the thermodynamic relations
\be
   u = \left(\partial f \over \partial \ln\Gamma\right)_{r_s},
\qquad
   \beta P/n = \frac13 \left[ u - 
   \left(\partial f \over \partial \ln r_s\right)_{\Gamma} \right].
\label{deriv}
\ee

The bottom line of the table is given for reference 
and presents the difference between the fit (\ref{fition}) and
numerical HNC data in the OCP model (i.e., without
the $ie$ contribution). 

\begin{figure}
    \begin{center}
    \leavevmode
    \epsfysize=100mm
    \epsfbox[70 150 550 675]{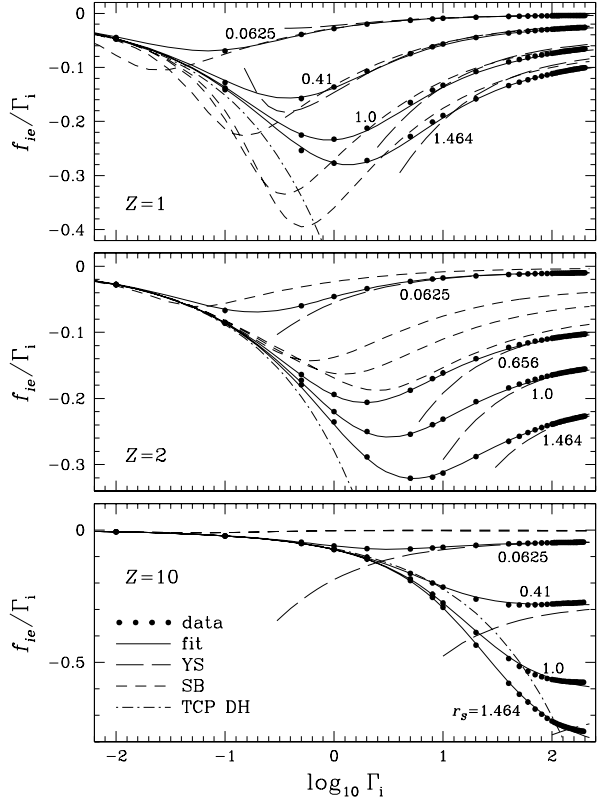}
    \end{center}
\caption{
Calculated (filled circles) and fitted (solid line)
$f_{ie}$ for $Z=1$, 2, and 10, 
for several indicated values of $r_s$,
compared with 
the approximations of DH (dot-dashed lines), 
YS (long dashes), and SB (short dashes).
}
\label{fig-ie}
\end{figure}

The calculated and fitted values of $f_{ie}$ are shown 
in Fig.~\ref{fig-z6} for $Z=6$ and in 
Fig.~\ref{fig-ie} for $Z=1,2$, and 10. For comparison, 
we have plotted the fit of 
Yakovlev and Shalybkov \cite{YS} (YS) to their 
relativistic calculations, carried out in the 
zero-temperature approximation
(justified at small $r_s$ and large $\Gamma_e$).
In Fig.~\ref{fig-ie} we have also plotted $f_{ie}$ given by
an analytic expression of
Ebeling et al.\cite{Ebeling-etal} reproduced by SB\cite{SB}. 
For the hydrogen plasma ($Z=1$) it reproduces
the Pad\'e approximations of Ref.\cite{EbelingRichert}. 
One can see that the fit of YS, in the range of its validity,
agrees with our results. On the contrary, the approximation 
of Refs.\cite{Ebeling-etal,SB} 
is clearly invalid in most cases.
It exhibits unphysical behavior
around $\Gamma_i\sim0.1$, predicting an enhancement of screening 
with decreasing $r_s$ (e.~g., for $Z=1$ and $\Gamma_i=0.1$ 
it gives larger $f_{ie}$ at $r_s=0.41$
than at $r_s=1.464$).
Moreover,
the extrapolation to $Z>1$, proposed in Ref.\cite{Ebeling-etal}, 
severely underestimates the screening effects. 

\begin{table}[ht]
\caption{Root-mean-square and 
maximum relative differences between the fit
and the HNC calculations
for $f_{ii}+f_{ie}$, $u_{ii}+u_{ie}$, and $P_{ii}+P_{ie}$;
bottom line correponds to the OCP model.
}
\label{tab2}
\begin{tabular}{rcccccc}
  $Z$ & \multicolumn{2}{c}{$(\delta f/f)$ (\%)} & 
 \multicolumn{2}{c}{$(\delta u/u)$ (\%)} & 
\multicolumn{2}{c}{$(\delta P/P)$ (\%)}\\ 
& rms & max & rms & max & rms & max\\ 
\noalign{\smallskip}
\hline
\noalign{\smallskip}
 1 & 0.6 & 1.9 & 0.9 & 1.8 & 1.2 & 4.5\\
 2 & 0.4 & 1.1 & 0.7 & 1.8 & 0.7 & 3.0\\
 3 & 0.4 & 0.8 & 0.9 & 2.3 & 0.7 & 1.8\\
 4 & 0.5 & 1.2 & 1.3 & 3.1 & 0.9 & 1.9\\
 5 & 0.6 & 1.5 & 1.6 & 3.9 & 1.2 & 2.4\\
 6 & 0.7 & 1.8 & 1.8 & 4.3 & 1.5 & 2.8\\
 7 & 0.6 & 1.7 & 1.9 & 4.6 & 1.7 & 3.5\\
 8 & 0.7 & 1.5 & 1.9 & 4.6 & 1.9 & 4.1\\
10 & 0.6 & 1.2 & 1.8 & 3.9 & 2.0 & 4.1\\
12 & 0.5 & 1.2 & 1.6 & 3.2 & 1.9 & 4.5\\
14 & 0.5 & 1.2 & 1.9 & 4.5 & 1.6 & 3.2\\
20 & 0.5 & 1.1 & 1.2 & 2.8 & 1.8 & 4.5\\
26 & 0.6 & 1.7 & 1.2 & 2.6 & 1.3 & 3.8\\
OCP& 0.6 & 0.7 & 0.6 & 0.8 & 0.6 & 0.8\\
\end{tabular}
\end{table}

\begin{figure}
    \begin{center}
    \leavevmode
    \epsfysize=90mm
    \epsfbox[80 160 550 670]{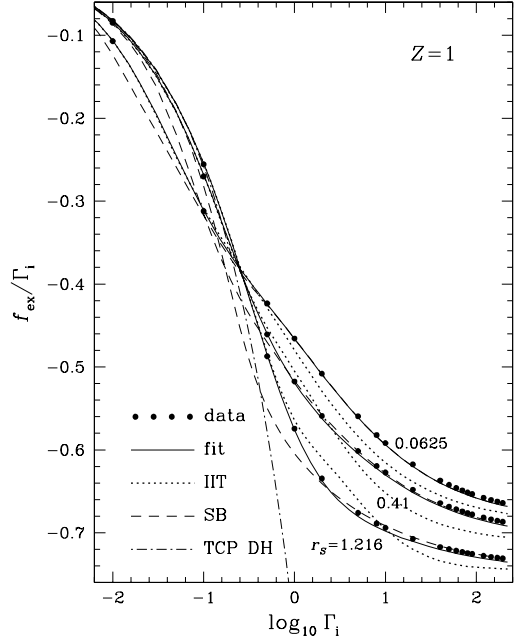}
    \end{center}
\caption{
Calculated (filled circles) and fitted (solid lines)
excess free energy,
 $f_{\rm ex}=x_i (f_{ii}+f_{ie})+x_e f_{ee}$, 
for $Z=1$ for different values of the density parameter $r_s$. The dot-dashed line is the DH formula while the
dashed and dotted lines
represent the approximations of 
SB\protect\cite{SB} and IIT\protect\cite{IIT},
respectively.
}
\label{fig-ex}
\end{figure}

Figure~\ref{fig-ex} exhibits an analogous comparison
for the excess free energy (\ref{f-ex}), for $Z=1$.
We have also plotted the 
Pad\'e-approximation of IIT \cite{IIT}. Although the disgressions 
between the fit and numerical 
data of Ref.\cite{IIT} lie within 0.4\%, 
there are significant deviations between the IIT fit and
our present results. This discrepancy originates from the 
relatively small number of numerical calculations used by IIT 
(32 computed values at $0.1\leq\Gamma_i\leq 10$ and
 $0.1\leq\theta\leq 10$). 
Our fit, 
based on a much larger set of numerical data,
not only reproduces these data but also
the numerical results of IIT \cite{IIT}.

\begin{figure}
    \begin{center}
    \leavevmode
    \epsfysize=90mm
    \epsfbox[70 150 550 685]{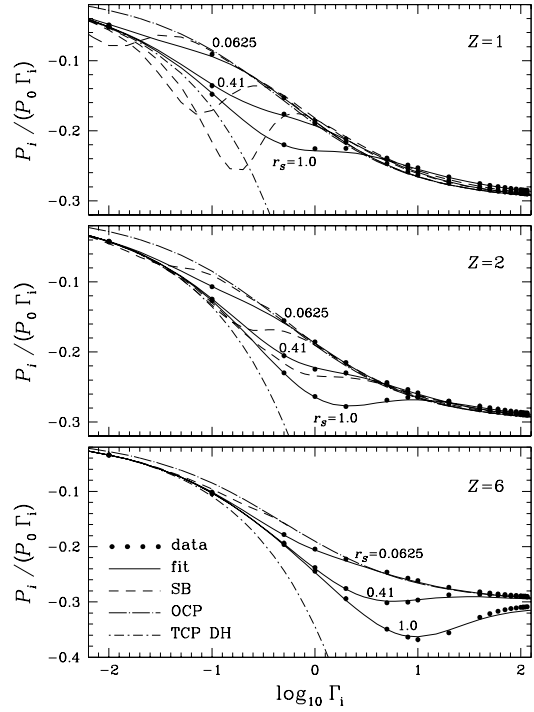}
    \end{center}
\caption{
Calculated (filled circles) and fitted (solid lines)
excess pressure of ions in the compressible
electron background, $P_i=P_{ii}+P_{ie}$,
in units of $P_0 \Gamma_i$, where $P_0=n_i k_B T$.
For comparison, the SB approximation
is shown by dashed lines,
and the DH and OCP approximations 
by dot-dashed lines.
}
\label{fig-pex}
\end{figure}

\begin{figure}
    \begin{center}
    \leavevmode
    \epsfysize=85mm
    \epsfbox[50 150 550 670]{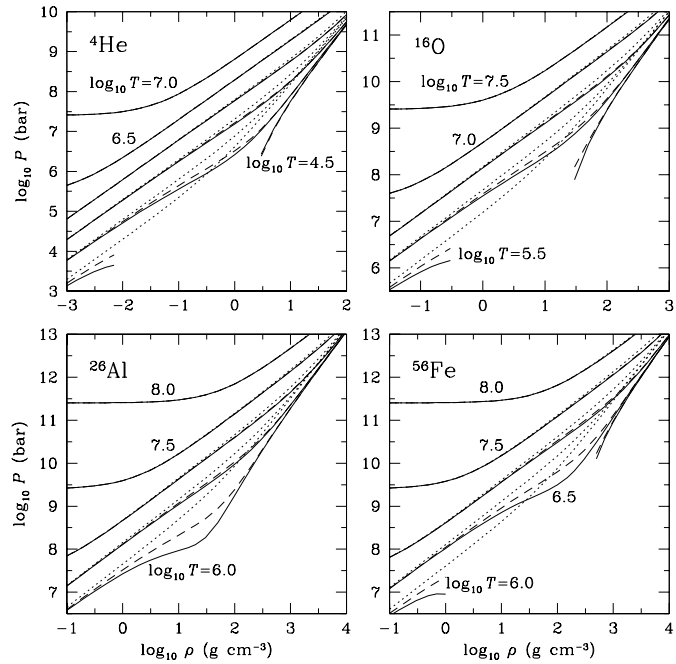}
    \end{center}
\caption{
Equations of state (EOS) of fully ionized plasmas
of four elements ($Z=4$, 8, 13, and 26) given by 
the present analytic approximations.
Solid lines show the pressure $P$ vs. density $\rho$
with account of the nonideality effects;
dots represent EOS of the perfect gas of ions and electrons;
dashes display an EOS in which the electron-ion 
screening effects are neglected.
The gaps in some isotherms
indicate the regions where the formation of
bound electron states can be expected.}
\label{fig-eos}
\end{figure}

The excess ionic pressure $P_{ii}+P_{ie}$
is shown in Fig.~\ref{fig-pex}.
Calculated data are compared with 
the pressure obtained by differentiation (\ref{deriv}) 
of our fit and of SB fit.
The DH approximation, shown for reference,
is calculated as the difference between the 
DH pressures
of the electron-ion TCP and the electron OCP.
The importance of the screening effects 
is verified by comparison of our calculated and fitted pressure
with the pressure of the OCP in the rigid background, $P_{ii}$,
also shown in the figure.

Figure \ref{fig-eos} demonstrates the validity of the EOS derived 
from our analytic formulae. The EOS of the perfect ion-electron gas
is compared with the EOS which includes the nonideality
of the electron and ion fluids
but neglects the $ie$ interactions;
solid lines show the complete EOS.
The gaps in some isotherms
indicate the regions where the formation of
bound states could not be neglected.
The significant deviations of the broken 
lines from the full lines
in certain ranges of $\rho$ and $T$
demonstrate the importance of the {\it ion-electron} 
screening effects.

\section{Multi-ionic mixtures}

The multi-ionic mixture is a straightforward generalization of the previous
single-ion model. In that case the effective Hamiltonian (\ref{heff}) reads :
\be 
H^{\rm eff}=K_i\,+\,\frac{1}{2V}\sum_{{\bf k} \ne 0}\frac{4\pi e^2}{k^2}
\left[
\frac {\rho_{Z\bf k}\rho_{Z\bf k}^\ast}{\epsilon(k)}
-N_i\langle Z^2 \rangle \right]
  \label{heff2}
\ee 
where  
$\rho_{Z\bf k}=\sum_i Z_i \rho_{i\bf k}$ are the Fourier components of the
ion charge number fluctuations.

For the binary ionic mixture in a rigid electron
background ($\epsilon(k)=1\, \forall k$), 
the excess (non-ideal) free energy of
the mixture, as well as the related thermodynamic quantities, 
can be expressed with 
high accuracy by the so-called ``linear mixing rule'' (LMR)
in terms of the free energy of the pure phases :
\be 
f_{\rm ex}(Z_1,Z_2,\Gamma_e,x_1) 
\approx x_1 f_{\rm ex}(\Gamma_1,x_1=1) + (1-x_1)f_{\rm ex}(\Gamma_2,x_1=0),
\label{freebim}
\ee
where $\Gamma_i= \Gamma_e Z_i^{5/3}$ and $x_1=N_1/(N_1+N_2)$. 
The very high level of accuracy 
of the LMR (\ref{freebim}) 
was first demonstrated by Hansen et al.\cite{HTV} 
and confirmed later on
by several authors, using very accurate MC 
calculations (e.g., \cite{DWSC,Ros}).

The validity of the LMR in the case of an ionic mixture immersed 
in a {\it responsive} finite-temperature electron
background, as described by the Hamiltonian (\ref{heff2}), 
has been examined by Hansen et al.\cite{HTV} in the 
first-order thermodynamic perturbation
approximation, and more recently
by Chabrier and Ashcroft \cite{ChabrierAshcroft},
who have solved the HNC equations with the effective screened potentials
for arbitrary charge ratios ranging from a symmetric case ($Z_2/Z_1\sim 1$)
to a highly asymmetric case ($Z_2/Z_1\gg 1$).
These authors found that the LMR remains valid to a high degree of accuracy
when the electron response is taken into account in the interionic potential,
except possibly for highly asymmetric mixtures in the region of
weak degeneracy of the electron gas (where the departure from linearity can
reach a few percent).

\section{Conclusions}
We have developed a completely analytic model for 
the free energy of fully ionized electron-ion Coulomb plasmas. 
The ideal part
of the free energy of electrons and ions
is described by Eqs.~(\ref{id_i})--(\ref{n_e}) and is accurately represented by
the analytic fits given by
Eqs.~(\ref{fitIa})--(\ref{chi}). 
Note that these formulae provide the thermodynamic 
quantities of a free electron gas for {\it any} 
degeneracy and relativity.
For the excess free energy 
of the electron fluid at finite temperature, we adopt
the analytic approximation from Ref.\cite{IIT}.
For the excess free energy of the classical ionic OCP,
we provide a simple interpolation (\ref{fition}) 
which accurately reproduces the Monte Carlo results at $\Gamma_i\ge 1$ 
and the Debye-H\"uckel-Abe limit for $\Gamma_i\ll 1$.
Finally, we have taken into account the ion-electron interactions
by solving the hypernetted-chain equations for a large set  of
the parameters $\Gamma_i$, $r_s$, and $Z$ 
and constructed an analytic fit given by Eq.~(\ref{fitscr}).
Our analytic formulae reproduce $f_{ii}+f_{ie}$
with accuracy $\sim 1-2$\%,
 and the derivatives of this function with respect to $r_s$ and $\Gamma_i$
 give an excess internal energy and pressure
with relative errors not larger than a few percent.
This analytic approximation 
is significantly more accurate than previous approximations
of the free energy of the electron-ion plasmas.

As mentioned in the introduction, 
our calculations imply full ionization, i.e.
point-like ions from which their bound electrons are stripped
completely.  
This model is realistic 
in various conditions at high temperatures or densities 
encountered in modern laser experiments and in various astrophysical 
situations like, for example, stellar, brown dwarf and 
giant planet interiors,
or the envelopes of neutron stars. In these situations, 
complete ionization can be safely assumed.
Furthermore, the present model can be used as the basis of
more elaborated equations of state aimed at describing 
the thermodynamic properties 
of partially ionized plasmas and ionization equilibrium. 
Work in this direction is in progress.

\begin{acknowledgements}
We thank D.~G. Yakovlev for useful
remarks on the manuscript.
A.Y.P. gratefully acknowlegdes generous hospitality 
and visiting professorship
in the theoretical astrophysics group
at the Ecole Normale Sup\'erieure de Lyon
and partial financial support from the grants
RFBR 96-02-16870a, DFG--RFBR 96-02-00177G, and INTAS 96-0542.
\end{acknowledgements}

\end{document}